\documentclass[10pt]{article}
\usepackage{mhchem}
\usepackage{fancyhdr}
\usepackage{extramarks} 
\usepackage{amsmath}
\usepackage{amsthm}
\usepackage{amsfonts}
\usepackage{siunitx}
\usepackage{tikz}
\usepackage[plain]{algorithm}
\usepackage{algpseudocode}
\usepackage{multirow}
\usepackage{booktabs}
\usepackage{palatino}
\usepackage{graphicx}
\usepackage{subfigure}
\usepackage[colorlinks,linkcolor=black,anchorcolor=black,citecolor=black,urlcolor=blue]{hyperref}
\usepackage{amsmath,bm}
\usepackage{booktabs}
\usepackage{mathtools}
\usepackage{amssymb}
\usepackage{caption}
\usepackage{capt-of}
\usepackage{mciteplus}
\usepackage{cite}
\usepackage{mathrsfs}
\usepackage{makecell}
\usepackage[title,titletoc,toc]{appendix}
\usepackage{xr}
\usepackage{parskip}
\usepackage{soul}
\usepackage{textcomp}
\usepackage[colaction]{multicol}
\usepackage[switch]{lineno}
\usepackage{lipsum}
\usepackage{etoolbox}
\usepackage{longtable}
\usepackage{array}
\usepackage{tablefootnote}
\usepackage{graphicx}
\usepackage{color, colortbl}
\usepackage{hhline}
\usepackage[T1]{fontenc} 
\usepackage{xr}

\definecolor{Lightblue}{rgb}{0.867,0.914,0.961}
\definecolor{Lightgreen}{rgb}{0.883,0.934,0.848}

\newcolumntype{C}[1]{>{\centering\arraybackslash}p{#1}}
\usetikzlibrary{automata,positioning}
\usetikzlibrary{automata,positioning}

\makeatletter
\newcommand*{\addFileDependency}[1]{
	\typeout{(#1)}
	\@addtofilelist{#1}
	\IfFileExists{#1}{}{\typeout{No file #1.}}
}
\makeatother

\newcommand*{\myexternaldocument}[1]{%
	\externaldocument{#1}%
	\addFileDependency{#1.tex}%
	\addFileDependency{#1.aux}%
}

\myexternaldocument{Supporting-information}

\begin{document}
	\title{Multiscale differential geometry learning for protein flexibility analysis}

\author{Hongsong Feng$^1$, Jeffrey Y. Zhao$^2$, and Guo-Wei Wei$^{1,3,4}$\footnote{
		Corresponding author.		Email: weig@msu.edu} \\
	\\
	$^1$ Department of Mathematics, \\
	Michigan State University, East Lansing, MI 48824, USA.\\
	$^2$ Vestavia Hills High School, Vestavia Hills, AL 35216, USA.\\
	$^3$ Department of Electrical and Computer Engineering,\\
	Michigan State University, East Lansing, MI 48824, USA. \\
	$^4$ Department of Biochemistry and Molecular Biology,\\
	Michigan State University, East Lansing, MI 48824, USA. \\
}
\date{\today} 

\maketitle
		
	\textbf{\large Abstract:}

Protein flexibility is crucial for understanding protein structures, functions, and dynamics, and it can be measured through experimental methods such as X-ray crystallography. Theoretical approaches have also been developed to predict B-factor values, which reflect protein flexibility. Previous models have made significant strides in analyzing B-factors by   fitting experimental data. In this study, we propose a novel approach for B-factor prediction using differential geometry theory, based on the assumption that the intrinsic properties of proteins reside on a family of low-dimensional manifolds embedded within the high-dimensional space of protein structures. By analyzing the mean and Gaussian curvatures of a set of kernel-function-defined low-dimensional manifolds, we develop effective and robust multiscale differential geometry (mDG) models. Our mDG model demonstrates a 27\% increase in accuracy compared to the classical Gaussian network model (GNM) in predicting B-factors for a dataset of 364 proteins. Additionally, by incorporating both global and local protein features, we construct a highly effective machine learning model for the blind prediction of B-factors. Extensive   least-squares approximations and machine learning-based blind predictions    validate the effectiveness of the mDG modeling approach for B-factor prediction.

	\pagenumbering{roman}
	\begin{verbatim}
	\end{verbatim}

	\newpage
	\clearpage
	\pagebreak
	%
	\newpage
	
	\setcounter{page}{1}
	\renewcommand{\thepage}{{\arabic{page}}}

\section{Introduction}\label{sec:introduction}

Proteins are polypeptide structures made up of one or more long chains of amino acid residues. According to the well-established sequence-structure-function dogma \cite{Anfinsen1973}, protein structures dictate their functions in various biological processes, including DNA replication, molecule transport, and providing structural support to cells. However, protein structures are not static; they exhibit fluctuations and thermodynamic movements under physiological conditions. These movements arise as proteins respond to external stimuli, and protein flexibility—measuring a protein's capacity to deform from its equilibrium state under external forces—represents an intrinsic property of the protein structure.

Protein flexibility can be assessed through experimental methods such as X-ray crystallography, nuclear magnetic resonance (NMR), and single-molecule force experiments. In X-ray crystallography, the Debye-Waller factor or beta factor (B-factor) characterizes protein flexibility by describing the attenuation of X-ray scattering due to thermal motion. Theoretically, the fluctuation amplitude of an atom in a protein correlates with its B-factor reported during the structure determination from X-ray diffraction data. However, reported B-factors may not fully account for variations in atomic diffraction cross-sections and chemical stability during data collection. NMR serves as another crucial technique for analyzing protein flexibility, allowing for the study of flexibility across different time scales and under physiological conditions.

Given that protein flexibility is linked to significant conformational variations, reactivity, and enzymatic functions \cite{teilum2009functional,teilum2011protein,ma2005usefulness}, analyzing protein flexibility is essential for understanding protein structure, function, and dynamics \cite{Frauenfelder1991}. Understanding protein flexibility is also important for docking \cite{chandrika2009managing} and computational drug design
\cite{carlson2000accommodating,teague2003implications}. This necessity has driven the development of numerous theoretical approaches to predict B-factors for specific protein structures, including molecular dynamics (MD) \cite{McCammon1977}, normal mode analysis (NMA) \cite{Brooks1983,Go1983,Levitt1985,Tasumi1982}, and elastic network models (ENM) \cite{Atilgan2001,Bahar1998,Bahar1997,Hinsen1998,Li2002,Tama2001}, as well as theories such as the Gaussian network model (GNM) \cite{Bahar1998,Bahar1997} and anisotropic network model (ANM) \cite{Atilgan2001} over the past few decades. While MD simulations can provide comprehensive conformational landscapes of proteins using an all-atom representation, they require extensive computational resources for long time integrations involving a significant number of degrees of freedom. To mitigate this issue, time-independent approaches are often utilized, functioning as time-harmonic approximations of Newton's equations \cite{Park2013}.

As one of the earliest time-independent B-factor prediction methods, the NMA \cite{Brooks1983,Go1983,Levitt1985,Tasumi1982} employs Hooke's Law to create an elastic mass-and-spring network for alpha carbons ($C_{\alpha}$). In this model, each $C_{\alpha}$ is represented as a node, with edges connecting nodes if their Euclidean distance is below a predefined threshold. This network effectively captures local covalent and non-covalent interactions between atoms and their neighbors and can be represented by a Hamiltonian interaction matrix. Eigenvalue analysis of this matrix yields the characteristic frequencies of the protein and predicts B-factors. The performance of elastic network computations has been further enhanced by various elastic network models (ENM) \cite{Atilgan2001,Bahar1998,Bahar1997,Hinsen1998,Li2002,Tama2001}. Notably, the anisotropic network model (ANM) \cite{Atilgan2001} functions as an NMA incorporating only the leading elasticity/MD potential and establishes connections between particles regardless of their chemical bonds \cite{flory1976statistical}. By disregarding anisotropic motion in the ANM network, the Gaussian network model (GNM) \cite{Bahar1998,Bahar1997} accelerates B-factor analysis, being approximately one order of magnitude more efficient than most other network approaches \cite{Yang2008Coarse}. However, eigenvalue analysis in these network approaches requires matrix diagonalization, which has a computational complexity of $\mathcal{O}(N^3)$, where $N$ is the number of atoms in the network. This becomes computationally expensive for larger proteins, highlighting the need for more efficient flexibility analysis methods. Graph method was reported for protein flexibility analysis\cite{jacobs2001protein}.  Sequence-based predictions and analysis of protein flexibility were proposed as well \cite{schlessinger2005protein,de2012predyflexy,vander2021medusa}. variosu machine learning approaches have also been developed for protein flexibility predictions \cite{vander2021medusa,masters2023deep,song2024accurate}. 
AlphaFold2 and deep learning were utilized to elucidate enzyme conformational flexibility 
\cite{casadevall2023alphafold2}. Recently, Xu et al have proposed method to employ both sequence information and structure information for predicting protein B-factors \cite{xu2024opus}. 

 Flexibility and rigidity index (FRI) methods \cite{xia2013multiscale,opron2014fast} have emerged as a matrix-decomposition-free approach for B-factor prediction. The fundamental assumption behind FRI methods is that protein flexibility and rigidity can be fully determined from the protein structure alone, eliminating the need to refer back to the protein interaction Hamiltonian, thereby bypassing costly matrix diagonalization. As a geometric graph approach, FRI methods construct a distance matrix using radial basis functions to nonlinearly scale atom-to-atom distances \cite{Xia2013Stochastic}. This allows for the evaluation of the rigidity index of each atom, which reflects the compactness of biomolecular packing. Consequently, the inverse of the rigidity index yields the flexibility index, correlating to the B-factor for each $C_{\alpha}$ atom \cite{xia2013multiscale,opron2014fast}. The original FRI \cite{xia2013multiscale} has a computational complexity of $\mathcal{O}(N^2)$, while the fast FRI (fFRI) \cite{opron2014fast} can be accelerated to $\mathcal{O}(N)$ using a cell lists algorithm. The parameter-free fFRI has demonstrated approximately 10\% greater accuracy than the GNM for a set of 365 proteins, all while being orders of magnitude faster \cite{opron2014fast}. To capture multiscale atomic interactions, a multiscale flexibility rigidity index (mFRI) method \cite{Opron2015Communication} has been developed, employing multiple radial basis functions or kernels with varying parameterizations, resulting in significant improvements in the accuracy of FRI B-factor predictions. Additionally, the anisotropic FRI (aFRI) model has been introduced for high-quality anisotropic motion analysis of biomolecules \cite{opron2014fast}.

Topological data analysis (TDA) methods have also been applied to protein flexibility analysis. In \cite{Bramer2020}, atom-specific persistent homology was constructed as a local atomic-level representation of a molecule using a global topological tool. The resulting atom-specific topological features were integrated with machine learning algorithms for B-factor predictions. Furthermore, evolutionary homology (EH) was introduced in \cite{Cang2020} based on a time evolution-based filtration and topological persistence. By coupling with dynamical systems or chaotic oscillators, the corresponding EH captures time-dependent topological invariants of macromolecules, making it applicable to protein flexibility analysis.
Pun et al. reported machine learning-based prediction of RNA flexibility using weighted persistent homology \cite{pun2020weighted}.  

Relevant to the present work are differential geometry (DG) approaches for the multiscale modeling of biomolecular systems \cite{wei2010differential}. Differential geometry, a branch of mathematics, employs advanced techniques from calculus to investigate curves and surfaces. Early biomolecular applications of DG methods focused on solvation analysis, as molecular surface modeling is crucial for understanding the geometric interactions between a solute protein and its surrounding solvent environment. Several differential geometry-based solvation models have been developed for molecular surface construction and solvation analysis \cite{wei2005molecular,bates2006minimal,bates2008minimal}. Notably, the first variational solute-solvent interface, known as the minimal molecular surface (MMS), was proposed in 2006 based on Laplace-Beltrami flow \cite{bates2006minimal,bates2008minimal}. This was subsequently coupled with Poisson-Boltzmann (PB) and Poisson-Nernst-Planck (PNP) models to create a family of differential geometry-based multiscale models for predicting solvation free energies \cite{Chen2010Differential,Chen2011Differential} and ion channel transport \cite{Chen2012Quantum,Wei2012Variational}.

Recently, advancements in DG approaches for geometric learning of biomolecular properties have been achieved in \cite{nguyen2019dg}, where critical chemical, physical, and biological information is encoded in element interactive manifolds extracted from a high-dimensional structural data space via a multiscale discrete-to-continuum density mapping. Low-dimensional DG representations, such as element interactive curvatures, can then be paired with robust machine learning algorithms for biomolecular modeling, leading to accurate predictions of molecular solvation free energy, protein-ligand binding affinity, and drug toxicity \cite{nguyen2019dg}. In \cite{Rana2022EISA}, molecular surface representations of element interactive manifolds were constructed as low-dimensional surface-based descriptors, resulting in a significant dimensional reduction for geometric learning. The resulting element interactive surface area (EISA) score facilitates an accurate machine learning algorithm for predicting protein-ligand binding affinity. 
Ricci curvature (FPRC)-based machine learning models have been proposed for for protein--ligand binding affinity prediction \cite{wee2021forman}. 
More recently, multiscale differential geometry learning has been effectively applied to single-cell RNA sequencing data analysis \cite{Feng2024Multiscale}. Curvature-based approach was developed for cell state analysis  in single-cell transcriptomic data
\cite{huynh2024topological}.

The goal of this work is to introduce a multiscale differential geometry (mDG) model for protein flexibility analysis. In the mDG model, a correlation function will be constructed  as in the FRI methods \cite{xia2013multiscale,opron2014fast,Opron2015Communication}, based on atomic interactions of $C_{\alpha}$ atoms, which has a complexity of $\mathcal{O}(N^2)$ or $\mathcal{O}(N)$ when the fast FRI algorithm \cite{opron2014fast} is adopted. Nevertheless, instead of directly calculating the rigidity or flexibility index, the correction function will be treated as a high-dimensional manifold in this work, which embeds information of all atomic interactions. Based on the principle of DG, low-dimensional atom-atom interactive manifolds will be extracted by using curvature analysis. Moreover, a multiscale modeling will be carried out so that the resulted family of Riemannian manifolds can capture atom-atom interactions at different scales. By analytically calculating the curvatures at $C_{\alpha}$ atom centers, a set of mDG features is generated. Following the convention of the protein flexibility analysis, the mDG features are integrated with a   regression algorithm for B-factor prediction, which ensures a fair comparison of other prediction approaches. 

The rest of the paper is organized as follows.  In section 2, the proposed mDG model will be presented. Details on manifold extraction and curvature evaluation will be offered. 
Section 3 is dedicated to the numerical results to demonstrate the performance of the proposed algorithm. A comparison with several existing prediction methods will also be considered. A summary and future plan will be discussed at the end of this paper.

\section{Theory and algorithm}\label{sec:method}

Our approximation of protein flexibility, or B-factors, is based on the protein's 3D structure. Protein structures encompass a wide range of characteristic length scales associated with various molecular interactions, including covalent bonds, hydrogen bonds, van der Waals forces, alpha helices, and beta sheets, among others \cite{Opron2015Communication}. These molecular interactions are closely linked to protein flexibility, making it essential to account for these multiscale interactions in our mathematical modeling.

In this study, we propose multiscale differential geometry (mDG) modeling to capture the multiscale collective motions of macromolecules. To characterize atomic interactions at varying distances, we employ multiple correlation kernels with different scaling parameters. The process begins with the introduction of atomic interaction manifolds, followed by the derivation of multiscale atomic interaction curvatures. These curvatures are utilized in the present study for B-factor approximation. The performance of our mDG methods in predicting B-factors is demonstrated in the \hyperref[sec]{Experiments} section.


\begin{figure}[ht]
	\centering
	\includegraphics[width=0.75\linewidth]{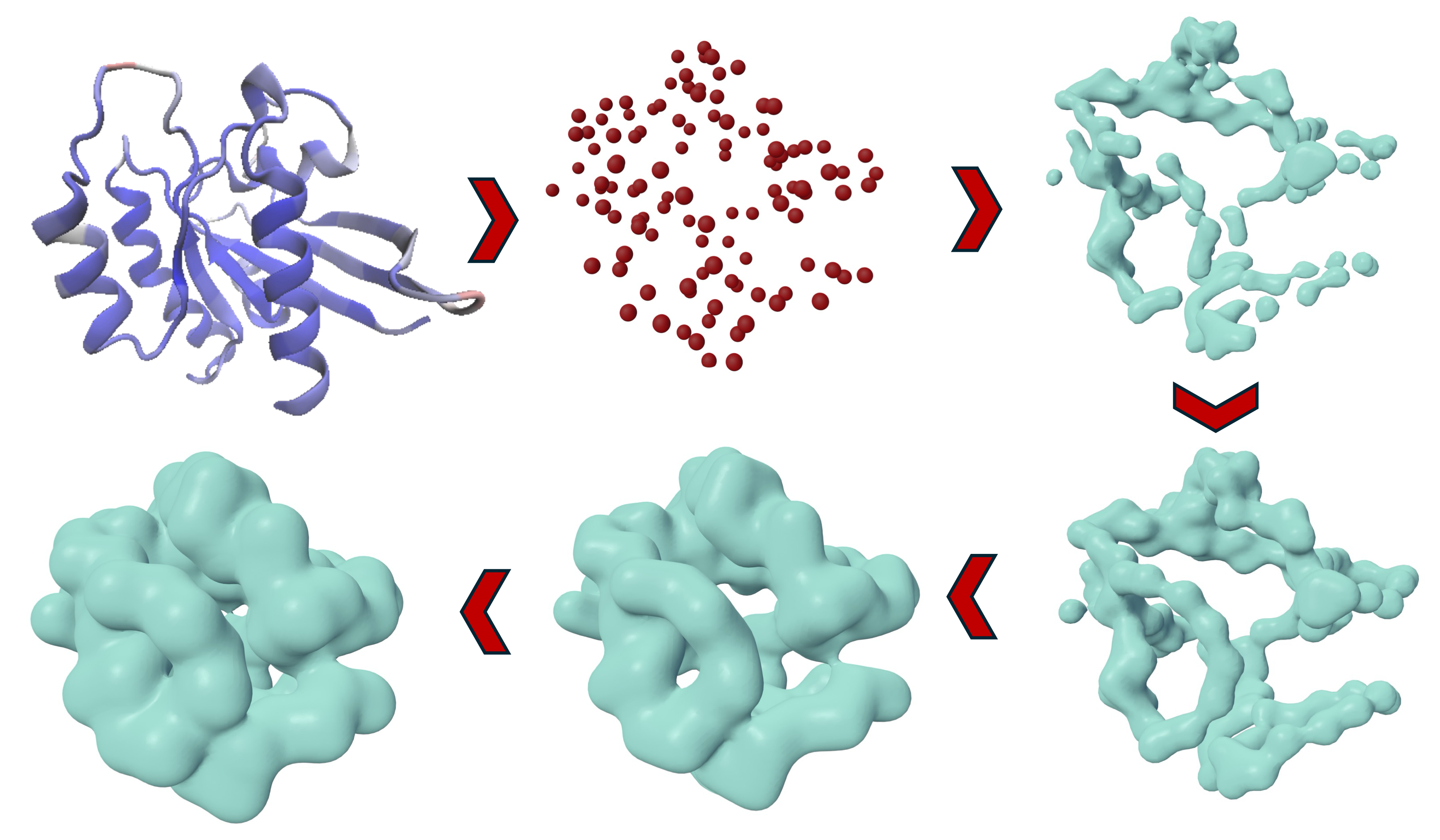} 
	\caption{{\footnotesize A series of manifold for the $C_{\alpha}$ atoms of protein 1CRR at 4 different isovalues with level set function \eqref{eq:atomic-interaction}. The density function is determined by kernel function \eqref{eq:generalized-exp} with parameter $\tau$=3 and $\kappa=5$. The red arrow indicates the isosurfaces generated with decreasing isovalues.}}
	\label{Fig:distribution}
\end{figure}

\subsection{Atomic interactive manifolds}
We propose using differential geometry modeling to capture the atomic interactions underlying atomic displacements or thermal motions. Our approach assumes that the intrinsic properties of proteins reside on a family of low-dimensional manifolds embedded within the high-dimensional space of protein structures. To achieve this, we convert discrete point cloud data (the atoms in a protein) into a continuous density distribution through a discrete-to-continuum mapping. The resulting density functions are then used to construct a series of low-dimensional manifolds that encapsulate these intrinsic atomic properties. Below, we present the construction of an atomic interaction manifold.

We follow the coarse-grained approach for B-factor approximation by solely considering $C_{\alpha}$ atoms in a protein. Assume a protein with the number of $C_{\alpha}$ atoms equal to $M$. Let $\mathcal{X} = \{ \mathbf{r}_1, ..., \mathbf{r}_M\}$ and vector $\mathbf{r}_i$ represents the 3D coordinate of $i$th $C_{\alpha}$ atom. Denote $\| \mathbf{r} - \mathbf{r}_i \|$ as the Euclidean distance between a point $\mathbf{r} \in \mathbb{R}^3$ and the atom $\mathbf{r}_i$. The unnormalized atomic interaction density can be given by a discrete to continuum mapping \cite{xia2013multiscale,opron2014fast,Opron2015Communication}
\begin{equation}\label{eq:atomic-interaction}
	\rho(\mathbf{r}; \{\eta_k\},\{w_k\}) = \sum_{j=1}^M w_j\Phi(\| \mathbf{r} - \mathbf{r}_j \|; \eta_j).
\end{equation}
Here, $\Phi$ is a correlation function with $C^2$ continunity and is chosen to have following admissibility properties:
\begin{align}\label{eq:admissibility}
	& \Phi(\| \mathbf{r} - \mathbf{r}_j \|; \eta_j) \to 0, \quad \text{as}\quad \|\mathbf{r} - \mathbf{r}_j\| \to \infty \\ 
	& \Phi(\| \mathbf{r} - \mathbf{r}_j \|; \eta_j) \to 1, \quad \text{as} \quad\|\mathbf{r} - \mathbf{r}_j\| \to 0.
\end{align}
The parameter $\eta_{j}$ is a characteristic distance between a point in the 3D space with the $j$th atom. Commonly used monotonically decaying kernel functions, such as radial basis functions, follow this pattern. Our previous work \cite{opron2014fast} has shown that the generalized exponential function,
\begin{equation} \label{eq:generalized-exp}
\Phi(\| \mathbf{r} - \mathbf{r}_j \|; \eta_j)  = e^{-(\| \mathbf{r} - \mathbf{r}_j \| / \eta_j)^\kappa}, \kappa > 0,
\end{equation} 
and generalized Lorentz function
\begin{equation} \label{eq:generalized-lor}
	\Phi(\| \mathbf{r} - \mathbf{r}_j \|; \eta_j)  = \frac{1}{1+(\| \mathbf{r} - \mathbf{r}_j \| / \eta_j)^v}, v > 0,
\end{equation} 
not only meet the admissibility assumptions, but also are excellent choices for protein flexibility analysis. 
For simplicity, we will consider only the generalized exponential function in this work. 

The kernel function \eqref{eq:generalized-exp}, which utilizes various resolution parameters $\eta$ and $\kappa$, characterizes the geometric and topological compactness of atomic interactions. A multiscale representation can be appropriately designed by selecting suitable values for these two resolution parameters. Our approach adheres to the FRI theory \cite{Opron2015Communication} for describing multiscale atomic interactions within molecules or for characterizing biomolecular interactions. Generally, a larger value of $\eta$ indicates a lower resolution and a slower decay, which has an equivalent effect to smaller power values of $\kappa$. In this work, we set $w_j^n=1$ as we focus our B-factor analysis on $C_{\alpha}$ atoms.

The correlation function $\rho$ in Eq. (\ref{eq:atomic-interaction}) for atomic interactions is governed by the scale parameter $\eta$ and the decay parameter $\kappa$. By selecting multiple values for $\eta$ and $\kappa$, we achieve a multiscale characterization of various atomic interactions. \autoref{Fig:distribution} illustrates the isosurfaces generated from a correlation function based on a single kernel function at different isovalues, constructed from a set of $C_{\alpha}$ atoms. Consequently, utilizing various kernels with different scaling configurations in the correlation function  \eqref{eq:atomic-interaction} enhances the embedding of different atomic interactions.


\subsection{Multiscale differential geometry of differential manifolds}
 
In the FRI methods \cite{xia2013multiscale,opron2014fast,Opron2015Communication}, the correlation function \eqref{eq:atomic-interaction} is used to directly define the rigidity and flexibility indices for B-factor prediction. A different usage of this function will be explored in this work. Mathematically, this function can be regarded as a manifold that encapsulates atomic interactions, making it feasible to use differential geometry to interpret these interactions. 
To this end, we first review the calculus on differentiable manifolds.

Let $\mathbf{M}: U \rightarrow \mathbb{R}^{n+1}$ be a $C^2$ mapping, where $U \subset \mathbb{R}^n$ is an open set with a compact closure \cite{bates2008minimal}. The mapping $\mathbf{M}(\mathbf{u}) = (M_1(\mathbf{u}), \ldots, M_n(\mathbf{u}), M_{n+1}(\mathbf{u}))$ represents a position vector on a hypersurface, where $\mathbf{u} = (u_1, \ldots, u_n) \in U$. The tangent or directional vectors of $\mathbf{M}$ are defined as $V_i = \frac{\partial \mathbf{M}}{\partial u_i}$ for $i = 1, \ldots, n$. The Jacobian matrix of $\mathbf{M}$ is given by $D\mathbf{M} = (V_1, V_2, \ldots, V_n)$. Using the notation $\langle \cdot \rangle$ for the Euclidean inner product in $\mathbb{R}^{n+1}$, the first fundamental form $I$ is defined as:
\[
I(V_i, V_j) := \langle V_i, V_j \rangle
\]
for any pair of tangent vectors $V_i, V_j \in T_{\mathbf{u}}\mathbf{M}$, where $T_{\mathbf{u}}\mathbf{M}$ denotes the tangent hyperplane at $\mathbf{M}(\mathbf{u})$. In the coordinates $\mathbf{M}(\mathbf{u})$, the first fundamental form can be expressed as a symmetric and positive definite matrix $(g_{ij}) = (I(V_i, V_j))$.

Let $\mathbf{N}(\mathbf{u})$ denote the unit normal vector defined by the Gauss map $\mathbf{N}: U \rightarrow \mathbb{R}^{n+1}$:
\[
\mathbf{N}(u_1, \ldots, u_n) = \frac{V_1 \times V_2 \times \ldots \times V_n}{\|V_1 \times V_2 \times \ldots \times V_n\|} \in \perp_{\mathbf{u}} \mathbf{M},
\]
where $\times$ represents the cross product in $\mathbb{R}^{n+1}$ and $\perp_{\mathbf{u}} \mathbf{M}$ is the normal space of $\mathbf{M}$ at the point $\mathbf{p} = \mathbf{M}(\mathbf{u})$. The normal vector $\mathbf{N}$ is orthogonal to the tangent hyperplane $T_{\mathbf{u}} \mathbf{M}$ at $\mathbf{M}(\mathbf{u})$. Using $\mathbf{N}$ and the tangent vectors $V_i$, the second fundamental form is defined as:
\[
II(V_i, V_j) = (h_{ij})_{i,j=1,\ldots,n} = \left(\left\langle - \frac{\partial \mathbf{N}}{\partial u_i}, V_j \right\rangle\right)_{ij}.
\]
The mean curvature $H$ is computed as $H = h_{ij} g^{ji}$, following the Einstein summation convention, with $g^{ji} = (g_{ij})^{-1}$. Additionally, the Gaussian curvature $K$ is given by:
\[
K = \frac{\text{Det}(h_{ij})}{\text{Det}(g_{ij})}.
\]

\subsection{Multiscale atomic interactive curvatures}

For the present study, it is sufficient to limit our discussion to the three-dimensional (3D) space ${\bf r}=(x,y,z)$, instead of general $\mathbb{R}^n$. Based on the kernel function $\rho$, different manifolds can be generated by considering different isovalues $\rho_0$ in the level set representation $\rho(x,y,z) = \rho_0$. Here, $\rho(x,y,z)$ can be assumed to be non-degenerate, i.e., the norm of its gradient is non-zero when it is equal to $\rho_0$. In the following discussion, 
we assume that its projection onto $z$ is non-zero. Then, a point on the iso-surface $\rho(x,y,z) = \rho_0$ has its $z$ coordinate being represented as $z=d(x,y)$, so that the iso-surface takes the form $\rho(x,y,d(x,y)) = \rho_0$. Note that if the projection onto $z$ is zero, we can carry out a similar process in $x$ or $y$ direction, and a similar conclusion holds.

Taking partial derivatives of $\rho(x,y,d(x,y)) = \rho_0$ with respect to $x$ and $y$ gives
\begin{align}\label{eq:dpho}
& \rho_x (x,y, d(x,y)) + \rho_z (x,y, d(x,y)) d_x = 0, \\
& \rho_y (x,y, d(x,y)) + \rho_z (x,y, d(x,y)) d_y = 0.
\end{align}
Consequently, we have $d_x = - \frac{\rho_x}{\rho_z}$ and $d_y = - \frac{\rho_y}{\rho_z}$. 
The coefficients in the first and second fundamental forms can then be defined as 
 \begin{align}\label{eq:coeffs}
E(x,y,d(x,y)) = \langle \rho_x, \rho_x \rangle, \quad F(x,y,d(x,y)) = \langle \rho_x, \rho_y \rangle, \quad
G(x,y,d(x,y)) = \langle \rho_y, \rho_y \rangle, \\
L(x,y,d(x,y)) = \langle \rho_{xx}, {\bf n} \rangle, \quad M(x,y,d(x,y)) = \langle \rho_{xy}, {\bf n} \rangle, \quad
N(x,y,d(x,y)) = \langle \rho_{yy}, {\bf n} \rangle,
\end{align}
where $\langle a, b \rangle$ denotes the inner product of $a$ and $b$, and ${\bf n}$ is the out normal direction of the iso-surface $\rho(x,y,d(x,y)) = \rho_0$.

In terms of these coefficients, the 3D Gaussian curvature $K$ and mean curvature $H$ can be computed as 
\begin{equation}\label{eq:K&H}
K = \frac{LN - M^2}{EG - F^2}, \quad 
H= \frac{1}{2} \frac{LG - 2 MF + NE}{EG - F^2}.
\end{equation}
Substituting $E$, $F$, $G$, $L$, $M$, $N$ into Eq. \eqref{eq:K&H}, the Gaussian curvature $K$ can be given as  \cite{xia2014multiscale}
\begin{align}\label{eq:gaussain_curvature}
	K =& \frac{2\rho_x \rho_y \rho_{xz} \rho_{yz} + 2 \rho_x \rho_z \rho_{xy} \rho_{yz} + 2 \rho_y \rho_z \rho_{xy} \rho_{xz}}{g^2}
	-\frac{2 \rho_x \rho_z \rho_{xz} \rho_{yy} + 2 \rho_y \rho_z \rho_{xx} \rho_{yz} + 2 \rho_x \rho_y \rho_{xy} \rho_{zz}}{g^2} \nonumber \\
	&+ \frac{\rho_z^2 \rho_{xx} \rho_{yy} + \rho_x^2 \rho_{yy} \rho_{zz} + \rho_y^2 \rho_{xx} \rho_{zz}}{g^2} - \frac{\rho_x^2 \rho_{yz}^2 + \rho_y^2 \rho_{xz}^2 + \rho_z^2 \rho_{xy}^2}{g^2},
\end{align}
where $g = \rho_x^2 + \rho_y^2 + \rho_z^2$. 
The mean curvature, which represents the average second derivative in the normal direction, is given by:
\begin{align}\label{eq:mean_curvature}
	H = -\frac{1}{2 g^{\frac{3}{2}}} \left[ 2 \rho_x \rho_y \rho_{xy} + 2 \rho_x \rho_z \rho_{xz} + 2 \rho_y \rho_z \rho_{yz}
	- (\rho_y^2 + \rho_z^2) \rho_{xx} - (\rho_x^2 + \rho_z^2) \rho_{yy} - (\rho_x^2 + \rho_y^2) \rho_{zz} \right].
\end{align}
Additionally, the minimum curvature $\mu_{\text{min}}$ and maximum curvature $\mu_{\text{max}}$ can be determined as:
\begin{align*}
	\mu_{\text{min}} = H - \sqrt{H^2 - K}, \\
	\mu_{\text{max}} = H + \sqrt{H^2 - K}.
\end{align*}
In differential geometry, various curvature measures describe the deviation of a geometric object from flatness, applicable to curves, surfaces, and higher-dimensional manifolds. Gaussian curvature and mean curvature are particularly useful for characterizing atomic interactions.

We note that  the Gaussian curvature and mean curvature computed in \eqref{eq:gaussain_curvature} and \eqref{eq:mean_curvature} are actually functions of the 3D space, i.e., $K(x,y,z)$ and $H(x,y,z)$ for any $(x,y,z) \in \mathbb{R}^3$, even though they are derived from iso-surface representation. 
Moreover, given the density function $\rho$, both the Gaussian curvature and mean curvature are continuous and can be computed analytically. 
This ensures that their expressions are free from numerical errors, making them well-suited for modeling atomic interactions. Additionally, the computational cost is relatively low since the density function includes only the $C_{\alpha}$ atoms, which are limited in number within the given datasets. In previous work \cite{opron2014fast}, fast algorithms were developed to leverage the rapid decay of the kernel effect within a narrow manifold band. This approach effectively addresses challenges related to the high computational demands for larger proteins in practical applications.

\subsection{Multiscale differential geometry for protein B-factor modeling}

Based on the interactive manifold described in \eqref{eq:atomic-interaction} and our differential geometry analysis, it is reasonable to use Gaussian and mean curvatures to provide a quantitative measure \((K, H)\) of an atom's interactions with others. To this end, we obtain a collection of Gaussian and mean curvatures as atomic features by varying the values of \(\eta\) and \(\kappa\) in the correlation function \eqref{eq:atomic-interaction}. Specifically, we consider a set of \(\eta\) values, \(\eta_i\) for \(i=1,2,\ldots,p\), and a set of \(\kappa\) values, \(\kappa_j\) for \(j=1,2,\ldots,q\). For the \(m\)th atom among \(M\) atoms in a given protein molecule, we define a curvature vector:
\begin{align}
C_m = \{(K_m^{i,j}, H_m^{i,j}) \mid i=1,2,\ldots,p; j=1,2,\ldots,q\}. \label{eq:features}
\end{align}
The above vector serves as a set of local features for a \(C_{\alpha}\) atom, representing its interactions with other atoms in the protein. While we focus on B-factor prediction for \(C_{\alpha}\) atoms in proteins, the approach presented in this work provides a general framework that can be used to predict B-factors for any atom in a protein.

In the current study, we consider two types of B-factor predictions. The first type follows the convention of protein flexibility analysis. We use the above mDG features to fit B-factors within a given protein using a   least squares minimization:
\begin{align}\label{eq:Bfactor}
	\min_{a^{i,j},b^{i,j}} \left\{ \sum_{m} \left| \sum_{i,j} a^{i,j} K_m^{i,j} + \sum_{i,j} b^{i,j} H_m^{i,j} + c - B_m^{e} \right|^2 \right\},
\end{align}
where ${B_m^e}$ are the experimental B-factors. The parameters ${a^{i,j}}$, ${b^{i,j}}$, and ${c}$ are to be determined through the optimization problem stated in \autoref{eq:Bfactor}. Note that the curvatures $K_m^{i,j}$ and $H_m^{i,j}$ are associated with the parameters $\eta_i$ and $\kappa_j$, which are preset in our multiscale modeling. Specifically, we use $\kappa = 2$ and $5$, and $\eta$ is set to $5, 9, 13, 17, 21, 25$, and $29$. This way, we can have B factor approximations for a set of atoms in a given protein and use certain metric to evaluate the approximation accuracy. For instance, we use Pearson correlation coefficient as detailed in the following section.


\subsection{Additional features for machine learning}

The second type of B-factor prediction we considered is a blind prediction for protein B-factors. We utilize mDG features as local descriptors of protein structures. These mDG features are combined with additional global and local protein features to build machine learning models. Each PDB structure includes a set of global features, such as the R-value, protein resolution, and the number of heavy atoms, which are provided in the PDB files. Local features for each protein include packing density, amino acid type, occupancy, and secondary structure information generated by the STRIDE software \cite{heinig2004stride}. STRIDE provides detailed secondary structure information for a protein based on its atomic coordinates from a PDB file, classifying each atom into categories such as alpha helix, 3-10 helix, $\pi$-helix, extended conformation, isolated bridge, turn, or coil. Additionally, STRIDE provides $\phi$ and $\psi$ angles and residue solvent-accessible area, resulting in a total of 12 secondary features. The packing density of each $C_{\alpha}$ atom in a protein is determined by the density of surrounding atoms. We defined short, medium, and long-range packing density features for each $C_{\alpha}$ atom. The packing density of the $i$th $C_{\alpha}$ atom is defined as
\begin{align}\
p_{i}^d =\frac{N_d}{N},
\end{align}	
where $d$ represents the specified cutoff distance in angstroms, $N_d$ denotes the number of atoms within the Euclidean distance $d$ from the $i$th atom, and $N$ is the total number of heavy atoms in the protein. The packing density cutoff values used in this study are provided in \autoref{table:packing-density}.
\begin{table}[htb!]
	\centering
	\begin{tabular}{c | c | c  }
		\hline
		\hline
		Short &  Medium & Long\\
		\hline
		$d<3$ & $3\ge d<5$ & $5\le d$\\
		\hline
		\hline
	\end{tabular}
	\caption{Packing density parameter in distance $d \AA$.}
	\label{table:packing-density}
\end{table}
Our mDG features, combined with the global and local features inherent in each PDB file, provide a comprehensive set of features for each $C_{\alpha}$ atom in the protein. For the blind predictions, we integrate these features with machine learning algorithms to build regression models. To demonstrate the performance of our machine learning model for blind predictions, we conduct two validation tasks: 10-fold cross-validation and leave-one-(protein)-out validation. The modeling and predictions focus on the B-factors of $C_{\alpha}$ atoms. Details and results are presented in the following section.

\begin{figure}[ht]
	\centering
	\includegraphics[width=0.75\linewidth]{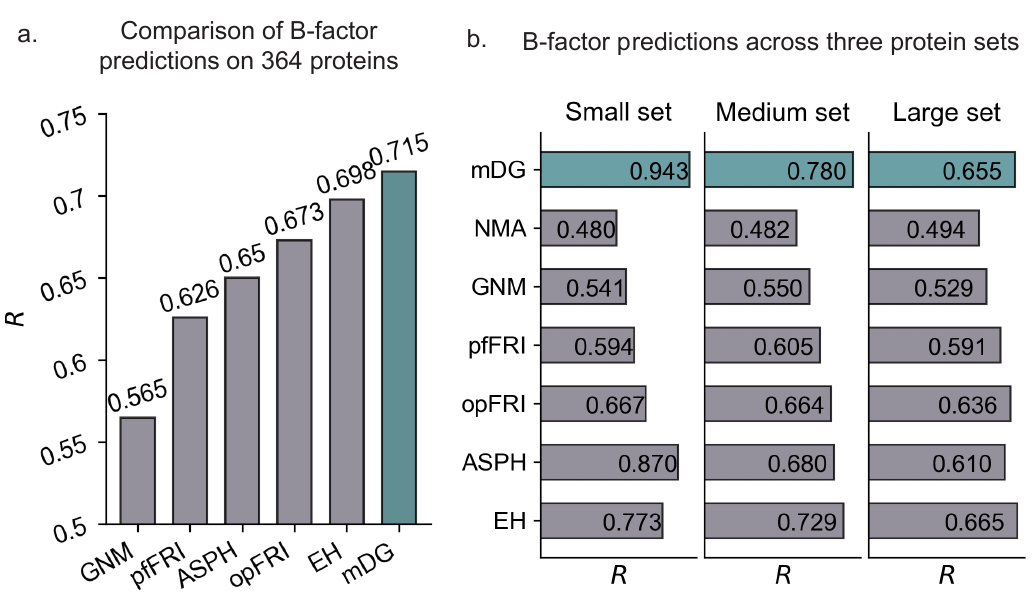}
	\caption{The average Pearson correlation coefficient (PCC) values using various advanced models for four B-factor prediction datasets are illustrated. (a) compares the average Pearson correlation coefficients between our mDG model and previous B-factor prediction models for 364 proteins. (b) compares the average PCC between our mDG model and previous B-factor prediction models across small, medium, and large protein datasets.}
	\label{Fig:bfactor-barplot}
\end{figure}

\section{Results}\label{sec:experiments}

\subsection{Data sets}

In this work, we use two datasets, one from Refs.~\cite{opron2014fast,Opron2015Communication} and the other from ~\cite{park2013coarse}. The first contains 364 proteins~\cite{opron2014fast,Opron2015Communication}, and the second \cite{park2013coarse} contains  three sets of proteins with small, medium, and large sizes, which are subsets of the 364 proteins. 


In the blind predictions, proteins 1ob4, 1ob7, 2oxl, and 3md5 are excluded from the data set because the STRIDE software is unable to provide features for these proteins. We exclude protein 1agn due to the known problems with this protein data. Proteins 1nko, 2oct, and 3fva are also excluded because these proteins have residues with B-factors reported as zero, which is unphysical. The following proteins were also excluded due to inconsistent protein data processed with STRIDE compared to original PDB data: 3dwv, 3mgn, 4dpz, 2j32, 3mea, 3a0m, 3ivv, 3w4q, 3p6j, and 2dko.

\subsection{Evaluation metrics}
To quantitatively assess our method for B-factor prediction, we use Pearson correlation coefficient (PCC):
\begin{align*}
	\text{PCC}({\bf{x}},{\bf{y}})=\frac{\sum_{m=1}^{M}(B_m^e-\bar{B}^e)(B_m^t-\bar{B}^t)}{\sqrt{\sum_{m=1}^{M} (B_m-\bar{B}^e)^2\sum_{m=1}^{M} (B_m^t-\bar{B}^t)^2}},
\end{align*}  
where $B_m^t, m = 1,2,\cdots,N$ are the predicted B-factors using the and $B_m^e, m = 1,2,\cdots,N$ are the experimental B-factors from the PDB file. Here $\bar{B}^e$ and $\bar{B}^t$ are the averaged B-factors.

\subsection{Experiments}

\subsubsection{Least square approximations}
For the least square approximations in B-factor  modeling, we consider the above four B-factor datasets. With the same benchmark datasets, there exist other advanced models in the literature, including Gauss network model (GNM)\cite{Bahar1998,Bahar1997}, flexibility rigidity index-based approaches such as pfFRI \cite{opron2014fast} and opFRI \cite{opron2014fast}, topology-based methods like atom-specific persistent homology (ASPH) \cite{bramer2020atom} and evolutionary homology (EH) \cite{cang2020evolutionary}. \autoref{Fig:bfactor-barplot}a gives the comparison between our mDG model and these approaches  in modeling the superset. Our model gives the highest average PCC value of 0.715 for the 364 proteins, surpassing the average PCC values of 0.698 for EH, 0.673 for opFRI, 0.65 for ASPH, 0.626 for pfFRI, and 0.565 for GNM, respectively. This represents a significant improvement of $2.4\%, 6.24\%, 10\%, 14.2\%,$ and $26.5\%$, respectively. \autoref{tab:364} presents the detailed comparative results between our mDG method and other approaches in terms of PCC value. Remarkably, mDG achieves higher PCC values than all these previous methods on 209 out of the 364 proteins.

Another three datasets consists for proteins of small, medium, and large sizes. We use these datasets to further validate the performance of our model. \autoref{Fig:bfactor-barplot}b compares the average PCC values of several models for each dataset. Normal mode analysis (NMA) \cite{Brooks1983} is another B-factor prediction model. mDG achieved average correlation coefficients of 0.943, 0.780, and 0.655 for the small, medium, and large protein sets, respectively. Our mDG model outperforms the previous methods, demonstrating improvements of 22\% and 7\% on the small and medium protein sets, respectively. Our mDG has average PCC value of 0.655 for the large protein dataset, which is only slightly below the previous state-of-the-art method EH \cite{cang2020evolutionary} with PCC of 0.665. These comprehensive comparisons demonstrate the robustness of mDG model for B-factor prediction across different protein sizes. In fact, the performance of mDG model for the dataset of large proteins can be improved by increasing the number of kernels, particularly by employing additional exponential kernel functions with larger $\eta$ values. A large protein has more atomic interactions. Embedding such kernel functions is beneficial to capturing those atomic interactions far away from the given atom.

The performance of GNM and NMA are poor across these four datasets with average PCC values less than 0.6. Overall, our mDG model significantly outperforms the two well-known models. Previous studies \cite{Opron2015Communication} have found that GNM and NMA fail to give reliable B-factor predictions on some proteins with hinge region or other special protein structures as shown in following case studies. There are various reasons for their low performance, partially due to the cutoff distance in building their models. Atomic interactions outside a cutoff distance are not accounted. Our multiscale differential geometry approach uses kernel functions to capture all atomic interactions within a molecule, while various scale distances, $\eta$, in the kernel functions allow to capture multi-resolutions atomic interactions. The multiscale different geometry modeling can properly address those challenges GNM and NMA face.

\subsubsection{Machine learning blind predictions}

For the blind predictions, our mDG features, along with other global and local features, are integrated with two types of machine learning algorithms: gradient boosting decision trees (GBDT) and random forest trees (RF). We conducted several experiments, the first of which involved leave-one-(protein)-out prediction using the four datasets mentioned above. We trained models ten times independently with different random seeds and calculated the average Pearson correlation coefficients from the ten sets of modeling predictions. The performance of the two types of machine learning models is shown in \autoref{table:Pearson-values-LOO}, where the GBDT-based models yield better predictions than the RF-based models.
\begin{table}[htb!]
	\centering
	\begin{tabular}{c | c c  c c }
		\hline
		\textbf{Protein set} &  \textbf{RF} & \textbf{GBDT}\\
		\hline
		Small & 0.460& 0.509\\
		Medium& 0.513&0.582\\
		Large & 0.475& 0.557\\
		Superset & 0.526 &0.587\\
		\hline
	\end{tabular}
	\caption{Average Pearson correlation coefficients (PCC) of leave-one (protein)-out predictions for the four B-factor datasets. The PCC results with random forest tree and gradient boosting decision tree modeling are compared.}
	\label{table:Pearson-values-LOO}
\end{table}

\begin{figure}[ht]
	\centering
	\includegraphics[width=0.7\linewidth]{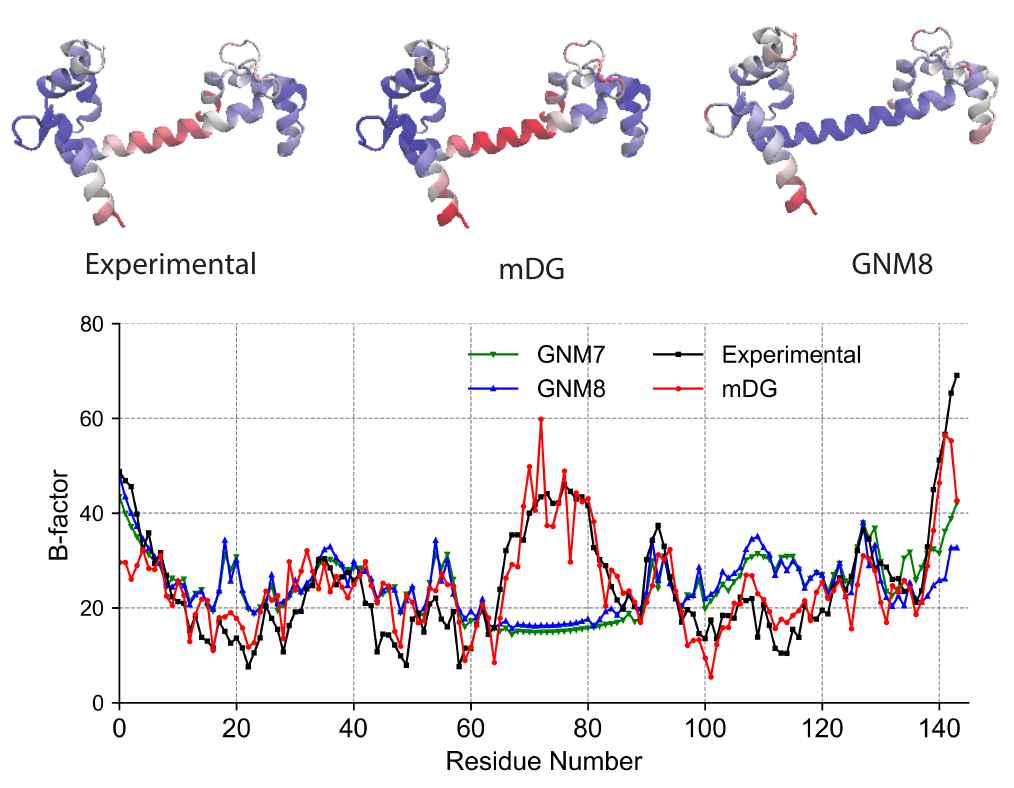} 
	\caption{{\footnotesize The upper section displays the 1CLL protein colored according to B-factor values from the experimental method, the mDG model, and the GNM model. The lower panel presents a detailed comparison of predicted B-factor values from various models alongside the experimental B-factor values. GNM7 and GNM8 refer to GNM modeling with cutoff distances of 7~\AA and 8~\AA, respectively.}}
	\label{Fig:Bfactors-compared-1CLLl}
\end{figure}
Least squares approximations were used to show the effectiveness of our mDG-based model. The overall performance of our model mDG is better than those literature models. Here, we present several case studies of relatively complex protein structures, which demonstrate the effectiveness of mDG modeling over others.

\begin{figure}[ht]
	\centering
	\includegraphics[width=0.7\linewidth]{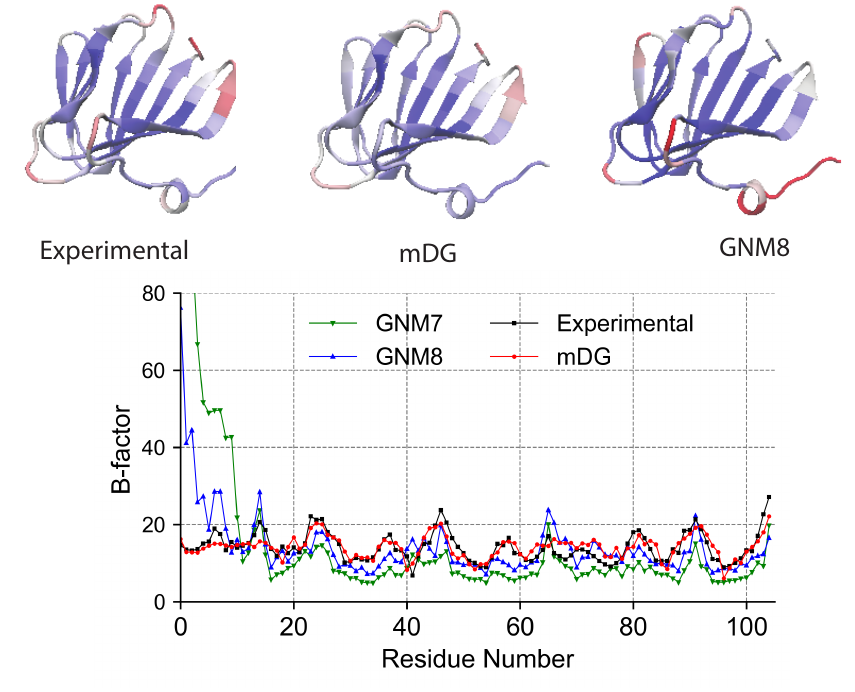} 
	\caption{{\footnotesize The upper section displays the 1V70 protein colored according to B-factor values from the experimental method, the mDG model, and the GNM model. The lower panel presents a detailed comparison of predicted B-factor values from various models alongside the experimental B-factor values. GNM7 and GNM8 refer to GNM modeling with cutoff distances of 7~\AA and 8~\AA, respectively.}}
	\label{Fig:Bfactors-compared-1V70}
\end{figure}

We also perform 10-fold cross-validation in our modeling. We use nine out of ten subsets of the 346 proteins to train our model, while the remaining subset is used for testing. Specifically, the features for atoms in the training proteins are combined together and used to train models. Those in the remaining proteins are used for testing. Ten different splitting were carried out. \autoref{table:Pearson-values-CV10-proteinwise} gives the average PCC values for two types of machine learning models. GBDT modeling give superior predictions than RF-based modeling. 

\begin{table}[htb!]
	\centering
	\begin{tabular}{c | c c  c c }
		\hline
		\textbf{Protein set} &  \textbf{RF} & \textbf{GBDT}\\
		\hline
		Superset & 0.842 & 0.859\\ 
		\hline
	\end{tabular}
	\caption{Average Pearson correlation coefficient (PCC) from 10-fold cross validation predictions with all atoms in the superset. The average PCC value is calculated from ten independent tests. The PCC results with random forest   and gradient boosting decision tree modeling are compared.}
	\label{table:Pearson-values-CV10-proteinwise}
\end{table}

We also performed an alternative 10-fold cross-validation for our modeling. The dataset consists of 346 proteins, containing over 74,000 atoms in total. In each of ten independent modeling, nine out of ten subsets of atoms are used for training the models, while the remaining subset is used for testing. \autoref{table:Pearson-values-CV10-proteinwise} gives the PCC values of two types of machine learning models with different algorithms. GBDT modeling yields slightly better predictions than RF-based modeling. 

\begin{table}[htb!]
	\centering
	\begin{tabular}{c | c c  c c }
		\hline
		\textbf{Protein set} &  \textbf{RF} & \textbf{GBDT}\\
		\hline
		Superset & 0.400 & 0.407\\ 
		\hline
	\end{tabular}
	\caption{Average Pearson correlation coefficient (PCC) from 10-fold cross validation predictions with all proteins in the superset. The average PCC value is calculated from ten independent tests. The PCC results with random forest tree and gradient boosting decision tree modeling are compared.}
	\label{table:Pearson-values-CV10-atomwise}
\end{table}

\begin{figure}[ht]
	\centering
	\includegraphics[width=0.7\linewidth]{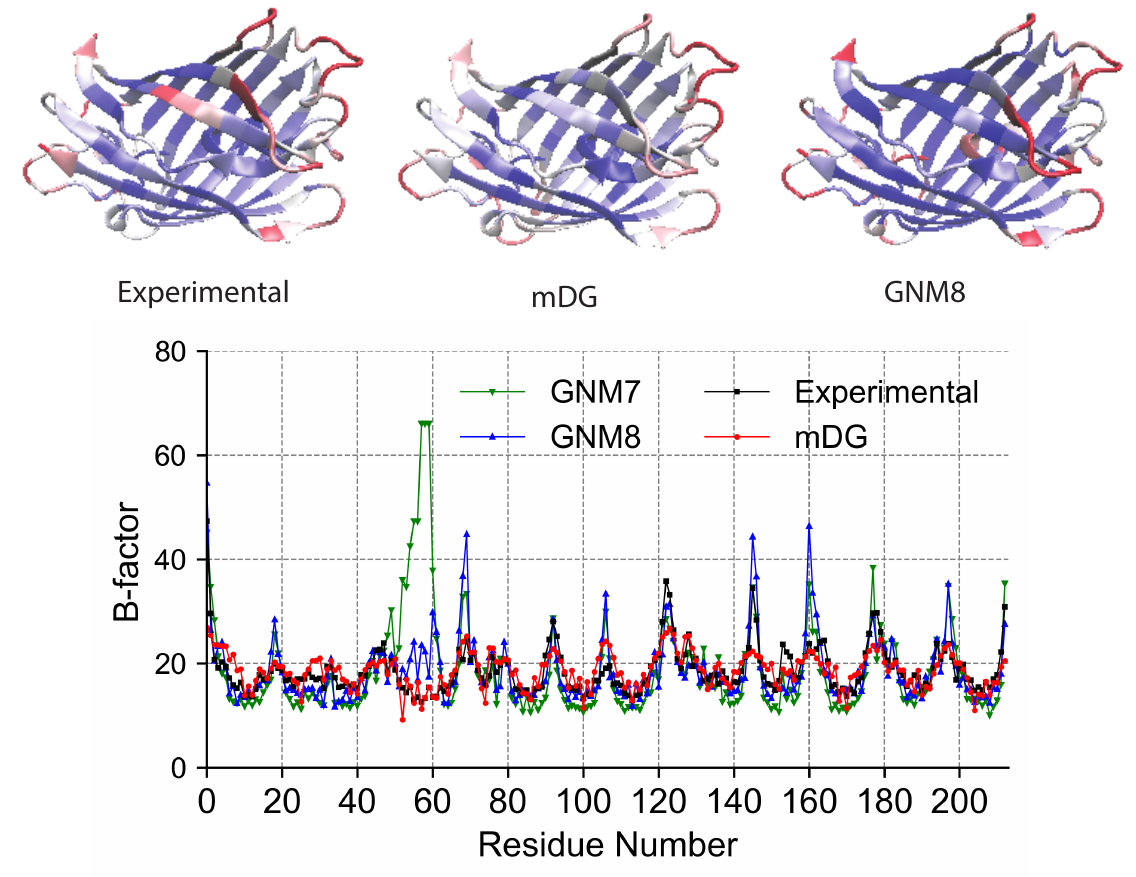} 
	\caption{{\footnotesize The upper section displays the 2HQK protein colored according to B-factor values from the experimental method, the mDG model, and the GNM model. The lower panel presents a detailed comparison of predicted B-factor values from various models alongside the experimental B-factor values. GNM7 and GNM8 refer to GNM modeling with cutoff distances of 7~\AA and 8~\AA, respectively.}}
	\label{Fig:Bfactors-compared-2HQK}
\end{figure}

\begin{figure}[ht]
	\centering
	\includegraphics[width=0.7\linewidth]{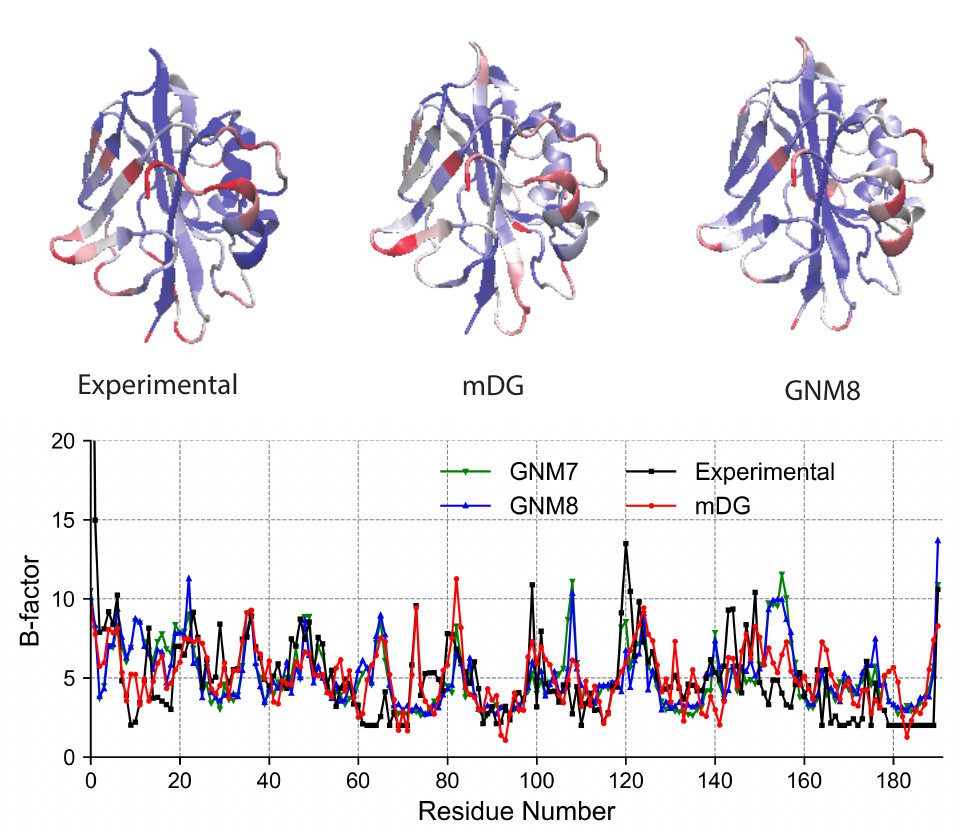}
	\caption{{\footnotesize The upper section displays the 2GZQ protein colored according to B-factor values from the experimental method, the mDG model, and the GNM model. The lower panel presents a detailed comparison of predicted B-factor values from various models alongside the experimental B-factor values. GNM7 and GNM8 refer to GNM modeling with cutoff distances of 7~\AA and 8~\AA, respectively.}}
	\label{Fig:Bfactors-compared-2GZQ}
\end{figure}

\subsubsection{Case studies of mDG modeling}


The first example is the protein calmodulin (PDBID: 1CLL) with 144 residues, which plays a key role in calcium signal transduction by modulating its interactions with various target proteins, such as kinases and phosphatases. Calmodulin's remarkable structural flexibility enables it to recognize a wide variety of target proteins. Proteins with hinge structures, like calmodulin, can undergo significant conformational changes, making it an excellent example for this type of analysis. The upper section in \autoref{Fig:Bfactors-compared-1CLLl} displays the proteins colored by the experimental or predicted B-factor values. The central region of calmodulin as shown in \autoref{Fig:Bfactors-compared-1CLLl} is a long $\alpha$-helix, which has large degree of flexibility based on B-factor values from the PDB 1CLL. Comparisons of B-factor predictions between the mDG and GNM models can be observed. It is clear that the B-factor values predicted by mDG are very close to the experimental values, while those predicted by GNM are less accurate. The lower section of \autoref{Fig:Bfactors-compared-1CLLl} presents a detailed numerical comparison, highlighting that the GNM method exhibits significant errors around residues 65-85. In contrast, the mDG method provides accurate B-factor predictions for these residues. GNM7 and GNM8 denote predictions made using the GNM model with cutoff distances of $7\AA$ and $8\AA$, respectively. Adjusting the cutoff distance in the GNM model does not resolve the inaccuracies observed in the hinge region.

The second example is a potential antibiotic synthesis protein (PDBID: 1V70) with 105 residues. Comparisons of the predicted B-factors using the mDG and GNM models are shown in \autoref{Fig:Bfactors-compared-1V70}. The B-factor coloring based on mDG predictions aligns closely with the experimental B-factor values, while the coloring based on the GNM method shows discrepancies. The problematic portion for B-factor prediction comes at one end of a protein chain. In this case, there is an overestimation of flexibility for residues 1-10 when using GNM. Again, GNM8 model gives marginally better predictions. Nevertheless, neither is able to reach the accuracy of mDG. As in the last case for 1CLL protein, 1V70 is moderate-size protein. mDG exhibit excellent B-factor predictions.

In the third example, we examine the flexibility prediction for the protein 2hqk, which has 232 residues. The lower section of \autoref{Fig:Bfactors-compared-2HQK} clearly shows that the GNM model exhibits significantly poor B-factor predictions around residues 50-60, with particularly pronounced errors at the recommended cutoff distance of 7~\AA. Even with an adjusted cutoff distance of 8~\AA, GNM8 does not resolve this issue. In contrast, the mDG model does not exhibit these problems. The protein coloring based on mDG predictions, shown in the upper section of \autoref{Fig:Bfactors-compared-2HQK}, closely resembles the experimental B-factors. Further inspection reveals that the problematic region of residues 50-60 corresponds to a small alpha-helical segment within the beta-barrel. This example highlights the sensitivity of the GNM model to cutoff distances and underscores how protein flexibility can be influenced by atomic interactions across various ranges. The mDG model, utilizing kernel functions with multiple scales, effectively captures these diverse molecular interactions, demonstrating its superiority over the GNM both theoretically and experimentally. Protein 2HQK has larger structure size than those in the above two cases. The numerical comparison validates the robustness of mDG in the B-factor prediction for a large-size protein.

As the final example, we consider protein 2GZQ, which has relatively low protein flexibility with B-factor values less than 15 except on one end of the protein chain. Overall, the predicted B-factor values using GNM models are slightly higher than those using mDG model. The predictions from our mDG model are closer to the experimental values than GNM predictions. \autoref{Fig:Bfactors-compared-2GZQ} gives detailed comparison between our mDG and GNM models. Protein 2GZQ is also having moderately large size with 203 residues. Our mDG model still remains effectiveness and yields accurate predictions of B-factor values for such large size proteins.

\section{Concluding remarks}

Protein flexibility is crucial to protein functions and its prediction is important for us to understand the protein properties. The intrinsic structural complexity hinders the understanding of protein flexibility. Effective computational approaches have been designed to predict B-factor values that reflect protein flexibility, such as GNM \cite{Bahar1998,Bahar1997}, pfFRI \cite{opron2014fast}, ASPH \cite{bramer2020atom}, opFRI \cite{opron2014fast}, EH \cite{cang2020evolutionary}, and NMA \cite{Brooks1983}. Our multiscale differential geometry model is a novel approach in this regard. Its effectiveness for B-factor predictions has been demonstrated with  least square approximation in comparison with these available methods. With the assumptions that atomic properties are sampled on low-dimensional manifolds in the high-dimensional protein structures, we construct a series of density-defined manifolds using soft-decaying kernel functions.  The mean and Gauss curvatures from differential geometry are proper tools for analyze atomic interactions. Different scales used in those density functions are beneficial to capturing atomic interactions in different distance ranges, which contributes to the effective multiscale modeling. In this sense, it is superior to the hard cutoff strategy in other methods such as GNM method, which may overlook the atomic interactions away from the cutoff distances. Our mDG method does not have such issue, especially by employing the multiscale strategy .

The integration of mDG features with additional global and local features intrinsic to protein structures and structure determination conditions  gives rise to useful machine learning models for blind predictions. Such blind-prediction models are useful to assess the B factor values or protein flexibility when the experimental B factors are unavailable. The extensive experiments with leave-one (protein)-out and 10-fold cross validations confirm the effectiveness and robustness of our machine learning models, especially the gradient boosting decision tree model.

\section{Acknowledgments}
This work was supported in part by NIH grants   R01AI164266  and R35GM148196, NSF grants DMS-2052983   and IIS-1900473,   MSU Foundation, and Bristol-Myers Squibb 65109.

\section{Appendix}

\autoref{tab:364} presents the comparisons of B-factor predictions from our mDG model and other literature models \cite{opron2014fast,park2013coarse}. The best predictions are highlighted in bold.

	{\footnotesize
	\begin{longtable}{|c|c|c|c|c|c||c|c|c|c|c|c|}
		\hline
		PDB ID & N & opFRI & pfFRI & GNM & mDG & PDB ID & N & opFRI & pfFRI & GNM & mDG \\
		\hline
		1ABA & 87 & 0.727 & 0.698 & 0.613 & \textbf{0.868} & 1AHO & 64 & 0.698 & 0.625 & 0.562 & \textbf{0.847} \\
		1AIE & 31 & 0.588 & 0.416 & 0.155 & \textbf{0.976} & 1AKG & 16 & 0.373 & 0.350 & 0.185 & \textbf{1.000} \\
		1ATG & 231 & \textbf{0.613} & 0.578 & 0.497 & 0.574 & 1BGF & 124 & \textbf{0.603} & 0.539 & 0.543 & 0.520 \\
		1BX7 & 51 & 0.726 & 0.623 & 0.706 & \textbf{0.802} & 1BYI & 224 & 0.543 & 0.491 & 0.552 & \textbf{0.568} \\
		1CCR & 111 & 0.580 & 0.512 & 0.351 & \textbf{0.767} & 1CYO & 88 & 0.751 & 0.702 & 0.741 & \textbf{0.823} \\
		1DF4 & 57 & \textbf{0.912} & 0.889 & 0.832 & 0.883 & 1E5K & 188 & 0.746 & 0.732 & \textbf{0.859} & 0.659 \\
		1ES5 & 260 & 0.653 & 0.638 & \textbf{0.677} & 0.661 & 1ETL & 12 & 0.710 & 0.609 & 0.628 & \textbf{1.000} \\
		1ETM & 12 & 0.544 & 0.393 & 0.432 & \textbf{1.000} & 1ETN & 12 & 0.089 & 0.023 & -0.274 & \textbf{1.000} \\
		1EW4 & 106 & 0.650 & 0.644 & 0.547 & \textbf{0.688} & 1F8R & 1932 & \textbf{0.878} & 0.859 & 0.738 & 0.635 \\
		1FF4 & 65 & 0.718 & 0.613 & 0.674 & \textbf{0.862} & 1FK5 & 93 & 0.590 & 0.568 & 0.485 & \textbf{0.661} \\
		1GCO & 1044 & \textbf{0.766} & 0.693 & 0.646 & 0.553 & 1GK7 & 39 & 0.845 & 0.773 & 0.821 & \textbf{0.935} \\
		1GVD & 52 & 0.781 & 0.732 & 0.591 & \textbf{0.875} & 1GXU & 88 & 0.748 & 0.634 & 0.421 & \textbf{0.833} \\
		1H6V & 2927 & \textbf{0.488} & 0.429 & 0.306 & 0.239 & 1HJE & 13 & 0.811 & 0.686 & 0.616 & \textbf{1.000} \\
		1I71 & 83 & 0.549 & 0.516 & 0.549 & \textbf{0.773} & 1IDP & 441 & \textbf{0.735} & 0.715 & 0.690 & 0.667 \\
		1IFR & 113 & 0.697 & 0.689 & 0.637 & \textbf{0.812} & 1K8U & 89 & 0.553 & 0.531 & 0.378 & \textbf{0.857} \\
		1KMM & 1499 & \textbf{0.749} & 0.744 & 0.558 & 0.488 & 1KNG & 144 & 0.547 & 0.536 & 0.512 & \textbf{0.652} \\
		1KR4 & 110 & 0.635 & 0.612 & 0.466 & \textbf{0.789} & 1KYC & 15 & 0.796 & 0.763 & 0.754 & \textbf{1.000} \\
		1LR7 & 73 & 0.679 & 0.657 & 0.620 & \textbf{0.783} & 1MF7 & 194 & 0.687 & 0.681 & \textbf{0.700} & 0.694 \\
		1N7E & 95 & 0.651 & 0.609 & 0.497 & \textbf{0.794} & 1NKD & 59 & 0.750 & 0.703 & 0.631 & \textbf{0.805} \\
		1NKO & 122 & 0.619 & 0.535 & 0.368 & \textbf{0.759} & 1NLS & 238 & \textbf{0.669} & 0.530 & 0.523 & 0.577 \\
		1NNX & 93 & 0.795 & 0.789 & 0.631 & \textbf{0.864} & 1NOA & 113 & 0.622 & 0.604 & 0.615 & \textbf{0.682} \\
		1NOT & 13 & 0.746 & 0.622 & 0.523 & \textbf{1.000} & 1O06 & 20 & 0.910 & 0.874 & 0.844 & \textbf{1.000} \\
		1O08 & 221 & \textbf{0.562} & 0.333 & 0.309 & 0.385 & 1OB4 & 16 & 0.776 & 0.763 & 0.750 & \textbf{1.000} \\
		1OB7 & 16 & 0.737 & 0.545 & 0.652 & \textbf{1.000} & 1OPD & 85 & 0.555 & 0.409 & 0.398 & \textbf{0.639} \\
		1P9I & 29 & 0.754 & 0.742 & 0.625 & \textbf{1.000} & 2CE0 & 99 & 0.706 & 0.598 & 0.529 & \textbf{0.871} \\
		2CG7 & 90 & 0.551 & 0.539 & 0.379 & \textbf{0.662} & 2COV & 534 & 0.846 & 0.823 & 0.812 & \textbf{0.850} \\
		2CWS & 227 & 0.647 & 0.640 & \textbf{0.696} & 0.537 & 2D5W & 1214 & \textbf{0.689} & 0.682 & 0.681 & 0.414 \\
		2DKO & 253 & \textbf{0.816} & 0.812 & 0.690 & 0.672 & 2DPL & 565 & 0.596 & 0.538 & \textbf{0.658} & 0.443 \\
		2DSX & 52 & 0.337 & 0.333 & 0.127 & \textbf{0.699} & 2E10 & 439 & \textbf{0.798} & 0.796 & 0.692 & 0.617 \\
		2E3H & 81 & \textbf{0.692} & 0.682 & 0.605 & 0.671 & 2EAQ & 89 & 0.753 & 0.690 & 0.695 & \textbf{0.866} \\
		2EHP & 248 & \textbf{0.804} & \textbf{0.804} & 0.773 & 0.711 & 2EHS & 75 & 0.720 & 0.713 & 0.747 & \textbf{0.777} \\
		2ERW & 53 & 0.461 & 0.253 & 0.199 & \textbf{0.899} & 2ETX & 389 & 0.580 & 0.556 & 0.632 & \textbf{0.653} \\
		2FB6 & 116 & 0.791 & 0.786 & 0.740 & \textbf{0.795} & 2FG1 & 157 & 0.620 & 0.617 & 0.584 & \textbf{0.719} \\
		2FN9 & 560 & 0.607 & 0.595 & \textbf{0.611} & 0.547 & 2FQ3 & 85 & 0.719 & 0.692 & 0.348 & \textbf{0.876} \\
		2G69 & 99 & 0.622 & 0.590 & 0.436 & \textbf{0.813} & 2G7O & 68 & 0.785 & 0.784 & 0.660 & \textbf{0.844} \\
		2G7S & 190 & \textbf{0.670} & 0.644 & 0.649 & 0.601 & 2GKG & 122 & 0.688 & 0.646 & 0.711 & \textbf{0.776} \\
		2GOM & 121 & 0.586 & 0.584 & 0.491 & \textbf{0.710} & 2GXG & 140 & \textbf{0.847} & 0.780 & 0.520 & 0.818 \\
		2GZQ & 191 & \textbf{0.505} & 0.382 & 0.369 & 0.480 & 2HQK & 213 & \textbf{0.824} & 0.809 & 0.365 & 0.738 \\
		2HYK & 238 & 0.585 & 0.575 & 0.510 & \textbf{0.619} & 2I24 & 113 & 0.593 & 0.498 & 0.494 & \textbf{0.614} \\
		2I49 & 398 & \textbf{0.714} & 0.683 & 0.601 & 0.671 & 2IBL & 108 & 0.629 & 0.625 & 0.352 & \textbf{0.700} \\
		2IGD & 61 & 0.585 & 0.481 & 0.386 & \textbf{0.824} & 2IMF & 203 & \textbf{0.652} & 0.625 & 0.514 & 0.574 \\
		2IP6 & 87 & 0.654 & 0.578 & 0.572 & \textbf{0.858} & 2IVY & 88 & 0.544 & 0.483 & 0.271 & \textbf{0.774} \\
		2J32 & 244 & \textbf{0.863} & 0.848 & 0.855 & 0.693 & 2J9W & 200 & \textbf{0.716} & 0.705 & 0.662 & 0.674 \\
		2JKU & 35 & 0.805 & 0.695 & 0.656 & \textbf{0.948} & 2JLI & 100 & 0.779 & 0.613 & 0.622 & \textbf{0.828} \\
		2JLJ & 115 & \textbf{0.741} & 0.720 & 0.527 & 0.635 & 2MCM & 113 & \textbf{0.789} & 0.713 & 0.639 & 0.649 \\
		2NLS & 36 & 0.605 & 0.559 & 0.530 & \textbf{0.900} & 2NR7 & 194 & \textbf{0.803} & 0.785 & 0.727 & 0.708 \\
		2NUH & 104 & 0.835 & 0.691 & 0.771 & \textbf{0.872} & 2O6X & 306 & \textbf{0.814} & 0.799 & 0.651 & 0.661 \\
		2OA2 & 132 & 0.571 & 0.456 & 0.458 & \textbf{0.582} & 2OCT & 192 & 0.567 & 0.550 & 0.540 & \textbf{0.569} \\
		2OHW & 256 & 0.614 & 0.539 & 0.475 & \textbf{0.754} & 2OKT & 342 & 0.433 & 0.411 & 0.336 & \textbf{0.500} \\
		2OL9 & 6 & 0.909 & 0.904 & 0.689 & \textbf{1.000} & 3BA1 & 312 & \textbf{0.661} & 0.624 & 0.621 & 0.516 \\
		3BED & 261 & \textbf{0.845} & 0.820 & 0.684 & 0.806 & 3BQX & 139 & 0.634 & 0.481 & 0.297 & \textbf{0.730} \\
		3BZQ & 99 & 0.532 & 0.516 & 0.466 & \textbf{0.751} & 3BZZ & 100 & 0.485 & 0.450 & 0.600 & \textbf{0.804} \\
		3DRF & 547 & \textbf{0.559} & 0.549 & 0.488 & 0.475 & 3DWV & 325 & \textbf{0.707} & 0.661 & 0.547 & 0.682 \\
		3E5T & 228 & 0.502 & 0.489 & 0.296 & \textbf{0.549} & 3E7R & 40 & 0.706 & 0.687 & 0.642 & \textbf{0.937} \\
		3EUR & 140 & 0.431 & 0.427 & \textbf{0.577} & 0.556 & 3F2Z & 149 & \textbf{0.824} & 0.792 & 0.740 & 0.786 \\
		3F7E & 254 & \textbf{0.812} & 0.803 & 0.811 & 0.750 & 3FCN & 158 & \textbf{0.640} & 0.606 & 0.632 & 0.436 \\
		3FE7 & 91 & 0.583 & 0.533 & 0.276 & \textbf{0.755} & 3FKE & 250 & 0.525 & 0.476 & 0.435 & \textbf{0.672} \\
		3FMY & 66 & 0.701 & 0.655 & 0.556 & \textbf{0.885} & 3FOD & 48 & 0.532 & 0.440 & -0.126 & \textbf{0.887} \\
		3FSO & 221 & \textbf{0.831} & 0.817 & 0.793 & 0.553 & 3FTD & 240 & \textbf{0.722} & 0.713 & 0.634 & 0.605 \\
		3FVA & 6 & 0.835 & 0.825 & 0.789 & \textbf{1.000} & 3G1S & 418 & 0.771 & 0.700 & 0.630 & \textbf{0.793} \\
		3GBW & 161 & 0.820 & 0.747 & 0.510 & \textbf{0.829} & 3GHJ & 116 & 0.732 & 0.511 & 0.196 & \textbf{0.828} \\
		3HFO & 197 & \textbf{0.691} & 0.670 & 0.518 & 0.569 & 3HHP & 1234 & \textbf{0.720} & 0.716 & 0.683 & 0.492 \\
		3HNY & 156 & \textbf{0.793} & 0.723 & 0.758 & 0.768 & 3HP4 & 183 & 0.534 & 0.500 & 0.573 & \textbf{0.653} \\
		3HWU & 144 & 0.754 & 0.748 & \textbf{0.841} & 0.675 & 3HYD & 7 & 0.966 & 0.950 & 0.867 & \textbf{1.000} \\
		3HZ8 & 192 & 0.617 & 0.502 & 0.475 & \textbf{0.729} & 3I2V & 124 & 0.486 & 0.441 & 0.301 & \textbf{0.642} \\
		3I2Z & 138 & 0.613 & 0.599 & 0.317 & \textbf{0.642} & 3I4O & 135 & 0.735 & 0.714 & 0.738 & \textbf{0.760} \\
		3I7M & 134 & 0.667 & 0.635 & 0.695 & \textbf{0.762} & 3IHS & 169 & 0.586 & 0.565 & 0.409 & \textbf{0.757} \\
		3IVV & 149 & \textbf{0.817} & 0.797 & 0.693 & 0.743 & 3K6Y & 227 & 0.586 & 0.535 & 0.301 & \textbf{0.695} \\
		3KBE & 140 & 0.705 & 0.704 & 0.611 & \textbf{0.773} & 3KGK & 190 & \textbf{0.784} & 0.775 & 0.680 & 0.754 \\
		3KZD & 85 & 0.647 & 0.611 & 0.475 & \textbf{0.811} & 3L41 & 220 & \textbf{0.718} & 0.716 & 0.669 & 0.636 \\
		3LAA & 169 & \textbf{0.827} & 0.647 & 0.659 & 0.575 & 3LAX & 106 & 0.734 & 0.730 & 0.584 & \textbf{0.757} \\
		3LG3 & 833 & \textbf{0.658} & 0.614 & 0.589 & 0.406 & 3LJI & 272 & \textbf{0.612} & 0.608 & 0.551 & 0.530 \\
		3M3P & 249 & \textbf{0.584} & 0.554 & 0.338 & 0.543 & 3M8J & 178 & \textbf{0.730} & 0.728 & 0.628 & 0.696 \\
		3M9J & 210 & 0.639 & 0.574 & 0.296 & \textbf{0.696} & 3M9Q & 176 & 0.591 & 0.510 & 0.471 & \textbf{0.625} \\
		3MAB & 173 & \textbf{0.664} & 0.591 & 0.451 & 0.610 & 3U6G & 248 & 0.635 & 0.632 & 0.526 & \textbf{0.656} \\
		3U97 & 77 & 0.753 & 0.736 & 0.712 & \textbf{0.762} & 3UCI & 72 & 0.589 & 0.526 & 0.495 & \textbf{0.624} \\
		3UR8 & 637 & \textbf{0.666} & 0.652 & 0.597 & 0.530 & 3US6 & 148 & \textbf{0.698} & 0.586 & 0.553 & 0.538 \\
		3V1A & 48 & 0.531 & 0.487 & 0.583 & \textbf{0.807} & 3V75 & 285 & 0.604 & 0.596 & 0.491 & \textbf{0.613} \\
		3VN0 & 193 & \textbf{0.840} & 0.837 & 0.812 & 0.836 & 3VOR & 182 & 0.602 & 0.557 & 0.484 & \textbf{0.690} \\
		3VUB & 101 & 0.625 & 0.610 & 0.607 & \textbf{0.739} & 3VVV & 108 & \textbf{0.833} & 0.741 & 0.753 & 0.736 \\
		3VZ9 & 163 & 0.785 & 0.749 & 0.695 & \textbf{0.799} & 3W4Q & 773 & \textbf{0.737} & 0.725 & 0.649 & 0.593 \\
		3ZBD & 213 & \textbf{0.651} & 0.516 & 0.632 & 0.649 & 3ZIT & 152 & 0.430 & 0.404 & 0.392 & \textbf{0.528} \\
		3ZRX & 221 & \textbf{0.590} & 0.562 & 0.391 & 0.588 & 3ZSL & 138 & 0.691 & 0.687 & 0.526 & \textbf{0.711} \\
		3ZZP & 74 & 0.524 & 0.460 & 0.448 & \textbf{0.779} & 3ZZY & 226 & \textbf{0.746} & 0.709 & 0.728 & 0.542 \\
		4A02 & 166 & 0.618 & 0.516 & 0.303 & \textbf{0.743} & 4ACJ & 167 & 0.748 & 0.746 & \textbf{0.759} & 0.726 \\
		4AE7 & 186 & 0.724 & 0.717 & 0.717 & \textbf{0.767} & 4AM1 & 345 & \textbf{0.674} & 0.619 & 0.460 & 0.653 \\
		4ANN & 176 & \textbf{0.551} & 0.536 & 0.470 & 0.518 & 4AVR & 188 & \textbf{0.680} & 0.605 & 0.650 & 0.599 \\
		4AXY & 54 & 0.700 & 0.623 & 0.720 & \textbf{0.881} & 4B6G & 558 & \textbf{0.765} & 0.756 & 0.669 & 0.567 \\
		4B9G & 292 & \textbf{0.844} & 0.816 & 0.763 & 0.590 & 4DD5 & 387 & \textbf{0.615} & 0.596 & 0.351 & 0.604 \\
		4DKN & 423 & \textbf{0.781} & 0.761 & 0.539 & 0.654 & 4DND & 95 & 0.763 & 0.750 & 0.582 & \textbf{0.765} \\
		4DPZ & 109 & 0.730 & 0.726 & 0.651 & \textbf{0.767} & 4DQ7 & 328 & \textbf{0.690} & 0.683 & 0.376 & 0.683 \\
		1PEF & 18 & 0.888 & 0.826 & 0.808 & \textbf{1.000} & 1PEN & 16 & 0.516 & 0.465 & 0.270 & \textbf{1.000} \\
		1PMY & 123 & 0.671 & 0.654 & 0.685 & \textbf{0.745} & 1PZ4 & 114 & 0.828 & 0.781 & \textbf{0.843} & 0.818 \\
		1Q9B & 43 & 0.746 & 0.726 & 0.656 & \textbf{0.917} & 1QAU & 112 & 0.678 & 0.672 & 0.620 & \textbf{0.848} \\
		1QKI & 3912 & \textbf{0.809} & 0.751 & 0.645 & 0.547 & 1QTO & 122 & 0.543 & 0.520 & 0.334 & \textbf{0.671} \\
		1R29 & 122 & 0.650 & 0.631 & 0.556 & \textbf{0.709} & 1R7J & 90 & 0.789 & 0.621 & 0.368 & \textbf{0.806} \\
		1RJU & 36 & 0.517 & 0.447 & 0.431 & \textbf{0.955} & 1RRO & 112 & 0.435 & 0.372 & \textbf{0.529} & 0.503 \\
		1SAU & 114 & 0.742 & 0.671 & 0.596 & \textbf{0.774} & 1TGR & 104 & 0.720 & 0.711 & 0.714 & \textbf{0.843} \\
		1TZV & 141 & 0.837 & 0.820 & \textbf{0.841} & 0.711 & 1U06 & 55 & 0.474 & 0.429 & 0.434 & \textbf{0.871} \\
		1U7I & 267 & \textbf{0.778} & 0.762 & 0.691 & 0.728 & 1U9C & 221 & 0.600 & 0.577 & 0.522 & \textbf{0.725} \\
		1UHA & 83 & 0.726 & 0.665 & 0.638 & \textbf{0.738} & 1UKU & 102 & 0.665 & 0.661 & \textbf{0.742} & 0.721 \\
		1ULR & 87 & 0.639 & 0.594 & 0.495 & \textbf{0.665} & 1UOY & 64 & 0.713 & 0.653 & 0.671 & \textbf{0.838} \\
		1USE & 40 & 0.438 & 0.146 & -0.142 & \textbf{0.936} & 1USM & 77 & \textbf{0.832} & 0.809 & 0.798 & 0.815 \\
		1UTG & 70 & 0.691 & 0.610 & 0.538 & \textbf{0.782} & 1V05 & 96 & 0.629 & 0.599 & 0.632 & \textbf{0.728} \\
		1V70 & 105 & 0.622 & 0.492 & 0.162 & \textbf{0.788} & 1VRZ & 21 & 0.792 & 0.695 & 0.677 & \textbf{1.000} \\
		1W2L & 97 & 0.691 & 0.564 & 0.397 & \textbf{0.735} & 1WBE & 204 & 0.591 & 0.577 & 0.549 & \textbf{0.598} \\
		1WHI & 122 & 0.601 & 0.539 & 0.270 & \textbf{0.734} & 1WLY & 322 & \textbf{0.695} & 0.679 & 0.666 & 0.597 \\
		1WPA & 107 & \textbf{0.634} & 0.577 & 0.417 & 0.587 & 1X3O & 80 & 0.600 & 0.559 & 0.654 & \textbf{0.844} \\
		1XY1 & 18 & 0.832 & 0.645 & 0.447 & \textbf{1.000} & 1XY2 & 8 & 0.619 & 0.570 & 0.562 & \textbf{1.000} \\
		1Y6X & 87 & 0.596 & 0.524 & 0.366 & \textbf{0.882} & 1YJO & 6 & 0.375 & 0.333 & 0.434 & \textbf{1.000} \\
		1YZM & 46 & 0.842 & 0.834 & \textbf{0.901} & 0.884 & 1Z21 & 96 & 0.662 & 0.638 & 0.433 & \textbf{0.718} \\
		1ZCE & 146 & \textbf{0.808} & 0.757 & 0.770 & 0.764 & 1ZVA & 75 & 0.756 & 0.579 & 0.690 & \textbf{0.796} \\
		2A50 & 457 & \textbf{0.564} & 0.524 & 0.281 & 0.561 & 2AGK & 233 & \textbf{0.705} & 0.694 & 0.512 & \textbf{0.705} \\
		2AH1 & 939 & \textbf{0.684} & 0.593 & 0.521 & 0.429 & 2B0A & 186 & 0.639 & 0.603 & 0.467 & \textbf{0.694} \\
		2BCM & 413 & \textbf{0.555} & 0.551 & 0.477 & 0.479 & 2BF9 & 36 & 0.606 & 0.554 & 0.680 & \textbf{0.951} \\
		2BRF & 100 & 0.795 & 0.764 & 0.710 & \textbf{0.877} & 2C71 & 205 & \textbf{0.658} & 0.649 & 0.560 & 0.585 \\
		2OLX & 4 & 0.917 & 0.888 & 0.885 & \textbf{1.000} & 2PKT & 93 & 0.162 & 0.003 & 0.193 & \textbf{0.744} \\
		2PLT & 99 & 0.508 & 0.484 & 0.509 & \textbf{0.708} & 2PMR & 76 & 0.693 & 0.682 & 0.619 & \textbf{0.801} \\
		2POF & 440 & 0.682 & 0.651 & 0.589 & \textbf{0.692} & 2PPN & 107 & 0.677 & 0.638 & 0.668 & \textbf{0.763} \\
		2PSF & 608 & 0.526 & 0.500 & \textbf{0.565} & 0.317 & 2PTH & 193 & \textbf{0.822} & 0.784 & 0.767 & 0.778 \\
		2Q4N & 153 & 0.711 & 0.667 & \textbf{0.740} & 0.698 & 2Q52 & 412 & \textbf{0.756} & 0.748 & 0.621 & 0.666 \\
		2QJL & 99 & 0.594 & 0.584 & 0.594 & \textbf{0.770} & 2R16 & 176 & 0.582 & 0.495 & 0.618 & \textbf{0.646} \\
		2R6Q & 138 & 0.603 & 0.540 & 0.529 & \textbf{0.659} & 2RB8 & 93 & 0.727 & 0.614 & 0.517 & \textbf{0.794} \\
		2RE2 & 238 & 0.652 & 0.613 & \textbf{0.673} & 0.498 & 2RFR & 154 & 0.693 & 0.671 & \textbf{0.753} & 0.727 \\
		2V9V & 135 & 0.555 & 0.548 & 0.528 & \textbf{0.623} & 2VE8 & 515 & \textbf{0.744} & 0.643 & 0.616 & 0.637 \\
		2VH7 & 94 & 0.775 & 0.726 & 0.596 & \textbf{0.845} & 2VIM & 104 & 0.413 & 0.393 & 0.212 & \textbf{0.599} \\
		2VPA & 204 & \textbf{0.763} & 0.755 & 0.576 & 0.610 & 2VQ4 & 106 & 0.680 & 0.679 & 0.555 & \textbf{0.754} \\
		2VY8 & 149 & \textbf{0.770} & 0.724 & 0.533 & 0.702 & 2VYO & 210 & 0.675 & 0.648 & \textbf{0.729} & 0.633 \\
		2W1V & 548 & \textbf{0.680} & \textbf{0.680} & 0.571 & 0.562 & 2W2A & 350 & \textbf{0.706} & 0.638 & 0.589 & 0.519 \\
		2W6A & 117 & \textbf{0.823} & 0.748 & 0.647 & 0.634 & 2WJ5 & 96 & 0.484 & 0.440 & 0.357 & \textbf{0.698} \\
		2WUJ & 100 & 0.739 & 0.598 & 0.598 & \textbf{0.804} & 2WW7 & 150 & 0.499 & 0.471 & 0.356 & \textbf{0.569} \\
		2WWE & 111 & 0.692 & 0.582 & 0.628 & \textbf{0.883} & 2X1Q & 240 & \textbf{0.534} & 0.478 & 0.443 & 0.516 \\
		2X25 & 168 & 0.632 & 0.598 & 0.403 & \textbf{0.643} & 2X3M & 166 & \textbf{0.744} & 0.717 & 0.655 & 0.690 \\
		2X5Y & 171 & 0.718 & 0.705 & 0.694 & \textbf{0.754} & 2X9Z & 262 & \textbf{0.583} & 0.578 & 0.574 & 0.492 \\
		2XHF & 310 & \textbf{0.606} & 0.591 & 0.569 & 0.517 & 2Y0T & 101 & 0.778 & 0.774 & 0.798 & \textbf{0.799} \\
		2Y72 & 170 & \textbf{0.780} & 0.754 & 0.766 & 0.712 & 2Y7L & 319 & \textbf{0.928} & 0.797 & 0.747 & 0.671 \\
		2Y9F & 149 & 0.771 & 0.762 & 0.664 & \textbf{0.787} & 2YLB & 400 & \textbf{0.807} & \textbf{0.807} & 0.675 & 0.629 \\
		2YNY & 315 & \textbf{0.813} & 0.804 & 0.706 & 0.721 & 2ZCM & 357 & 0.458 & 0.422 & 0.420 & \textbf{0.525} \\
		2ZU1 & 360 & \textbf{0.689} & 0.672 & 0.653 & 0.600 & 3A0M & 148 & \textbf{0.807} & 0.712 & 0.392 & 0.710 \\
		3A7L & 128 & 0.713 & 0.663 & \textbf{0.756} & 0.639 & 3AMC & 614 & \textbf{0.675} & 0.669 & 0.581 & 0.670 \\
		3AUB & 116 & 0.614 & 0.608 & 0.637 & \textbf{0.735} & 3B5O & 230 & 0.644 & 0.629 & 0.601 & \textbf{0.725} \\
		3MD4 & 12 & 0.860 & 0.781 & 0.914 & \textbf{1.000} & 3MD5 & 12 & 0.649 & 0.413 & -0.218 & \textbf{1.000} \\
		3MEA & 166 & 0.669 & 0.669 & 0.600 & \textbf{0.737} & 3MGN & 348 & 0.205 & 0.119 & 0.193 & \textbf{0.704} \\
		3MRE & 383 & \textbf{0.661} & 0.641 & 0.567 & 0.487 & 3N11 & 325 & \textbf{0.614} & 0.583 & 0.517 & 0.613 \\
		3NE0 & 208 & \textbf{0.706} & 0.645 & 0.659 & 0.627 & 3NGG & 94 & 0.696 & 0.689 & \textbf{0.719} & 0.674 \\
		3NPV & 495 & \textbf{0.702} & 0.653 & 0.677 & 0.476 & 3NVG & 6 & 0.721 & 0.617 & 0.597 & \textbf{1.000} \\
		3NZL & 73 & 0.627 & 0.583 & 0.506 & \textbf{0.812} & 3O0P & 194 & 0.727 & 0.706 & \textbf{0.734} & 0.629 \\
		3O5P & 128 & \textbf{0.734} & 0.698 & 0.630 & 0.642 & 3OBQ & 150 & 0.649 & 0.645 & \textbf{0.655} & 0.576 \\
		3OQY & 234 & \textbf{0.698} & 0.686 & 0.637 & 0.619 & 3P6J & 125 & 0.774 & 0.767 & 0.810 & \textbf{0.838} \\
		3PD7 & 188 & \textbf{0.770} & 0.723 & 0.589 & 0.670 & 3PES & 165 & 0.697 & 0.642 & 0.683 & \textbf{0.736} \\
		3PID & 387 & 0.537 & 0.531 & \textbf{0.642} & 0.378 & 3PIW & 154 & \textbf{0.758} & 0.744 & 0.717 & 0.615 \\
		3PKV & 221 & \textbf{0.625} & 0.597 & 0.568 & 0.614 & 3PSM & 94 & \textbf{0.876} & 0.790 & 0.745 & 0.870 \\
		3PTL & 289 & 0.543 & 0.541 & 0.468 & \textbf{0.596} & 3PVE & 347 & \textbf{0.718} & 0.667 & 0.568 & 0.635 \\
		3PZ9 & 357 & \textbf{0.709} & \textbf{0.709} & 0.678 & 0.602 & 3PZZ & 12 & 0.945 & 0.922 & 0.950 & \textbf{1.000} \\
		3Q2X & 6 & 0.922 & 0.904 & 0.866 & \textbf{1.000} & 3Q6L & 131 & 0.622 & 0.577 & 0.605 & \textbf{0.728} \\
		3QDS & 284 & \textbf{0.780} & 0.745 & 0.568 & 0.656 & 3QPA & 197 & \textbf{0.587} & 0.442 & 0.503 & 0.469 \\
		3R6D & 221 & \textbf{0.688} & 0.669 & 0.495 & 0.681 & 3R87 & 132 & 0.452 & 0.419 & 0.286 & \textbf{0.589} \\
		3RQ9 & 162 & 0.510 & 0.403 & 0.242 & \textbf{0.675} & 3RY0 & 128 & 0.616 & 0.606 & 0.470 & \textbf{0.761} \\
		3RZY & 139 & 0.800 & 0.784 & \textbf{0.849} & 0.828 & 3S0A & 119 & 0.562 & 0.524 & 0.526 & \textbf{0.657} \\
		3SD2 & 86 & 0.523 & 0.421 & 0.237 & \textbf{0.760} & 3SEB & 238 & 0.801 & 0.712 & \textbf{0.826} & 0.583 \\
		3SED & 124 & 0.709 & 0.658 & 0.712 & \textbf{0.819} & 3SO6 & 150 & 0.675 & 0.666 & 0.630 & \textbf{0.756} \\
		3SR3 & 637 & 0.619 & 0.611 & \textbf{0.624} & 0.395 & 3SUK & 248 & \textbf{0.644} & 0.633 & 0.567 & 0.618 \\
		3SZH & 697 & \textbf{0.817} & 0.815 & 0.697 & 0.657 & 3T0H & 208 & \textbf{0.808} & 0.775 & 0.694 & 0.793 \\
		3T3K & 122 & \textbf{0.796} & 0.748 & 0.735 & 0.685 & 3T47 & 141 & 0.592 & 0.527 & 0.447 & \textbf{0.740} \\
		3TDN & 357 & 0.458 & 0.419 & 0.240 & \textbf{0.627} & 3TOW & 152 & 0.578 & 0.556 & 0.571 & \textbf{0.782} \\
		3TUA & 210 & 0.665 & 0.658 & 0.588 & \textbf{0.706} & 3TYS & 75 & 0.853 & 0.800 & 0.791 & \textbf{0.896} \\
		4DT4 & 160 & \textbf{0.776} & 0.738 & 0.716 & 0.708 & 4EK3 & 287 & \textbf{0.680} & \textbf{0.680} & 0.674 & 0.608 \\
		4ERY & 318 & \textbf{0.740} & 0.701 & 0.688 & 0.642 & 4ES1 & 95 & 0.648 & 0.625 & 0.551 & \textbf{0.790} \\
		4EUG & 225 & 0.570 & 0.529 & 0.405 & \textbf{0.612} & 4F01 & 448 & 0.633 & 0.372 & \textbf{0.688} & 0.640 \\
		4F3J & 143 & 0.617 & 0.598 & 0.551 & \textbf{0.734} & 4FR9 & 141 & 0.671 & 0.655 & 0.501 & \textbf{0.673} \\
		4G14 & 15 & 0.467 & 0.323 & 0.356 & \textbf{1.000} & 4G2E & 151 & 0.760 & 0.755 & 0.758 & \textbf{0.767} \\
		4G5X & 550 & \textbf{0.786} & 0.754 & 0.743 & 0.551 & 4G6C & 658 & \textbf{0.591} & 0.590 & 0.528 & 0.497 \\
		4G7X & 194 & \textbf{0.688} & 0.587 & 0.624 & 0.660 & 4GA2 & 144 & \textbf{0.528} & 0.485 & 0.406 & 0.513 \\
		4GMQ & 92 & 0.678 & 0.628 & 0.550 & \textbf{0.762} & 4GS3 & 90 & 0.544 & 0.522 & 0.547 & \textbf{0.821} \\
		4H4J & 236 & \textbf{0.810} & 0.806 & 0.689 & 0.705 & 4H89 & 168 & \textbf{0.682} & 0.588 & 0.596 & 0.523 \\
		4HDE & 168 & \textbf{0.745} & 0.728 & 0.615 & 0.585 & 4HJP & 281 & \textbf{0.703} & 0.649 & 0.510 & 0.620 \\
		4HWM & 117 & 0.638 & 0.622 & 0.499 & \textbf{0.801} & 4IL7 & 85 & 0.446 & 0.404 & 0.316 & \textbf{0.732} \\
		4J11 & 357 & \textbf{0.620} & 0.562 & 0.401 & 0.587 & 4J5O & 220 & \textbf{0.793} & 0.757 & 0.777 & 0.580 \\
		4J5Q & 146 & 0.742 & 0.742 & 0.689 & \textbf{0.859} & 4J78 & 305 & 0.658 & 0.648 & 0.608 & \textbf{0.733} \\
		4JG2 & 185 & \textbf{0.746} & 0.736 & 0.543 & 0.661 & 4JVU & 207 & 0.723 & 0.697 & 0.553 & \textbf{0.736} \\
		4JYP & 534 & \textbf{0.688} & 0.682 & 0.538 & 0.573 & 4KEF & 133 & 0.580 & 0.530 & 0.324 & \textbf{0.705} \\
		5CYT & 103 & 0.441 & 0.421 & 0.331 & \textbf{0.641} & 6RXN & 45 & 0.614 & 0.574 & 0.594 & \textbf{0.889} \\
		\hline
		\end{longtable}
\label{tab:364}		
\captionof{table}{The comparison of correlation coefficient of mGLI with previous methods including opFRI, prFRI, and GNM. N refers to the number of residues in the protein. And the best value for each protein is marked in bold.}
}

\bibliographystyle{unsrt}

\end{document}